# Auto-Cal: Automated and Continuous Geo-Referencing of All-Sky Imagers Using Fisheye Lens Modeling and Star Tracks

Sudha Kapali[1], Michael P. Henderson[1], Juanita Riccobono[1], Michael A. Migliozzi[1], Robert B. Kerr[1]

[1] Computational Physics, Inc

## 1 Abstract

A fully automated and continuous calibration framework for All-Sky Imagers (ASIs) that significantly enhances the spatial accuracy and reliability of geo-referenced ASI data is presented. The technique addresses a critical bottleneck in ASI image data reliability and usability for real time space weather via automated geo-referencing under real-world field conditions. The system corrects the lens distortion in ASIs using a well-established fisheye lens model and automatically estimates camera orientation in terms of roll, pitch, and yaw angles relative to True North and the horizontal plane perpendicular to the zenith using star tracking. Unlike traditional methods that require manual intervention and periodic recalibration, Auto-Cal performs nightly unattended recalibrations using observed stellar motion, adapting to mechanical shifts or environmental changes. Each calibration step includes formal error estimates, allowing end users to assess the confidence of geo-located data in real-time. This capability enables dependable ASI operations in remote or unmanned settings and supports higher-fidelity integration with other geophysical instruments. Auto-Cal provides a scalable foundation for maintaining a large array of ASIs, thus enabling long-term atmospheric monitoring and real-time space weather alerts.

## 2 Introduction

All-Sky Imagers (ASIs) are essential tools for capturing large-scale atmospheric phenomena, including auroras, airglow, traveling ionospheric disturbances (TIDs) and mesoscale traveling ionospheric disturbances (MSTIDs). However, the scientific usefulness of ASI data heavily depends on accurate spatial calibration. Current spatial calibration methods rely on an expert to detect that the instrument requires calibration or recalibration, involve human input in the method, and are time-consuming, thus delaying the availability of calibrated image data to the end user. To address these limitations, this work introduces an automated and continuous calibration technique that provides trustworthy ASI image data with error estimates with each night of data acquisition.

Our solution utilizes a distinctive two-step calibration process. The first step is conducted in the laboratory during instrument assembly to determine the ASI's intrinsic parameters – specifically optical center, focal lengths along the x- and y- axes, and the distortion model of the fisheye lens system. These parameters are stable and, once determined, do not require recalibration in the field unless the instrument undergoes a significant repair or upgrade. The second step, which is the calibration of the camera's orientation with respect to True North (or South) and a horizontal plane perpendicular to the zenith, takes place during field deployment. These parameters – the roll, pitch and yaw angles relative to True North and the horizontal plane perpendicular to the zenith - are sensitive to gradual environmental changes or sudden instrument position shifts. Ideally, the instrument should be calibrated continuously to derive these parameters to deliver trustworthy ASI image data to the end user. Our technique is responsive to the requirement by automating the process deriving the camera orientation using star tracks in the ASI imagery.

This paper makes the following key contributions:

1. Fully automated calibration system: The paper presents a novel system that automates both lens distortion correction and camera pose estimation (roll, pitch, yaw angles) using stellar observation, thus eliminating the need for manual intervention for estimating the camera orientation.

2. Continuous nightly calibration: The system recalibrates automatically each night, adapts to environmental or physical changes (e.g., bumps or shifts in orientation), and includes quality metrics for tracking calibration accuracy over time.

3. Availability of calibration error estimates for quality assessment: Each nightly calibration includes quantitative error estimates derived from deviation between expected and observed star positions allowing users to assess how trustworthy the geo-referenced image data is.

# 3 Background and Motivation

## 3.1 Background on ASIs and Space Weather

Ground-based All-Sky Imagers (ASIs) are a cornerstone of space weather and auroral research, providing continuous, wide-field views of ionospheric phenomena such as auroral arcs, diffuse precipitation, and traveling ionospheric disturbances [Figure 1]. All-Sky Imagers are designed to capture airglow and auroral emissions from the upper atmosphere, and each emission corresponds to a specific altitude range and physical process. The ground footprint refers to the area on the Earth's surface from which light (emissions) observed by an All-Sky Imager appears to originate, as projected downward from the emission altitude [Figure 2].

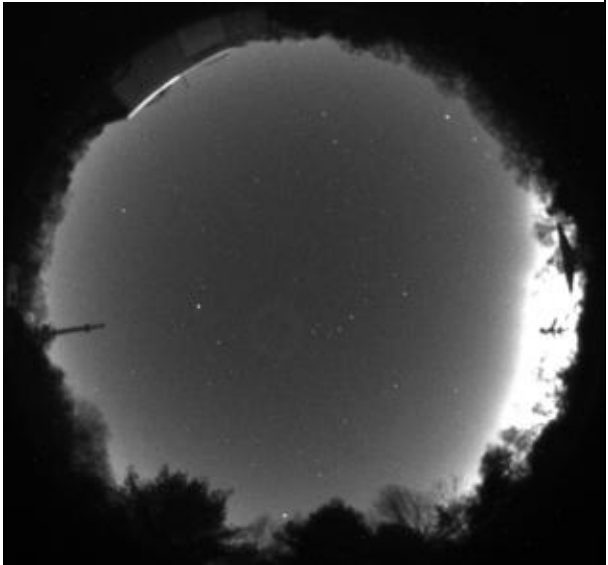

Figure 1: Image acquired from CPI's ASI at MIT Haystack Observatory on 04/18/2025 at 4:23:38 UT, while a geomagnetic storm was in progress.

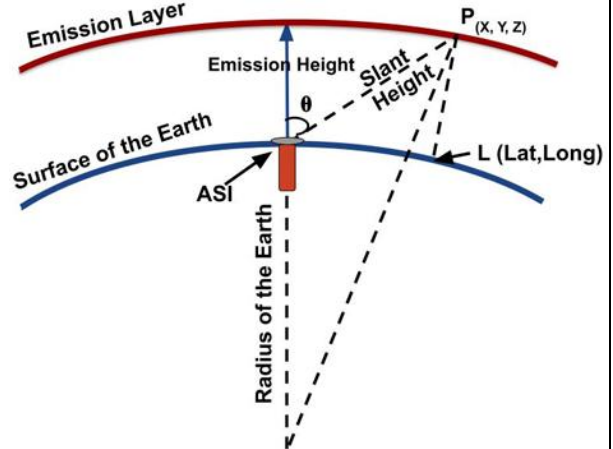

Figure 2: A schematic representation of the ground footprint at geographic location L in latitude and longitude coordinates of a point P(x, y, z) on the emission layer. Point P is at an assumed emission height and makes an angle $\theta$ with the optical axis

The pixel locations in these ASI images must be accurately mapped to geographic coordinates. This calibration step is essential for tracking the motion of geophysical features over time for real time space weather alerts as well as for integrating ASI data with measurements by other instruments, such as radars or satellites.

Previous studies have demonstrated the importance of accurate geo-referencing in ASI-based investigations. For example, Colpitts *et al.* (2013) performed a detailed comparison of discrete auroral features using simultaneous observations from THEMIS ASIs and the FAST satellite. Their study required high-confidence spatial and temporal alignment of ASI observations with satellite trajectories, illustrating the need for precise ASI calibration when comparing multi-instrument measurements. Similarly, Gillies *et al.* (2014) conducted a statistical survey of quiet auroral arc orientations and their dependence on interplanetary magnetic field (IMF) conditions. Their analysis relied on consistently calibrated ASI observations across multiple sites and events to resolve longitudinal and latitudinal arc orientations—highlighting the need for stable and repeatable image-to-geographic coordinate mappings.

The scientific demand for real-time high-resolution, geo-referenced All-Sky Imager data is growing, especially for applications in space weather monitoring such as auroral boundary detection. A broad spectrum of users—including power utilities, researchers, and the public— benefit from accurate real-time geo-referenced ASI data.

There are two parts to solving the problem of geo-referencing ASI imagery. Firstly, the wide field of view of the ASI (close to 180°) is achieved with a lens system that includes a fisheye lens [Figure 3]. This introduces a non-linearity in the image which is more apparent for objects that are at wider angles from the optical axis. The fisheye lens expands the field of view in the ASI image data[Figure 1, Figure 4], projecting objects located at wider angles from the optical axis toward the edges of the image, where they appear radially compressed. In contrast, objects closer to the optical axis are spaced farther apart radially. The georeferencing solution must account for this effect in its formulation.

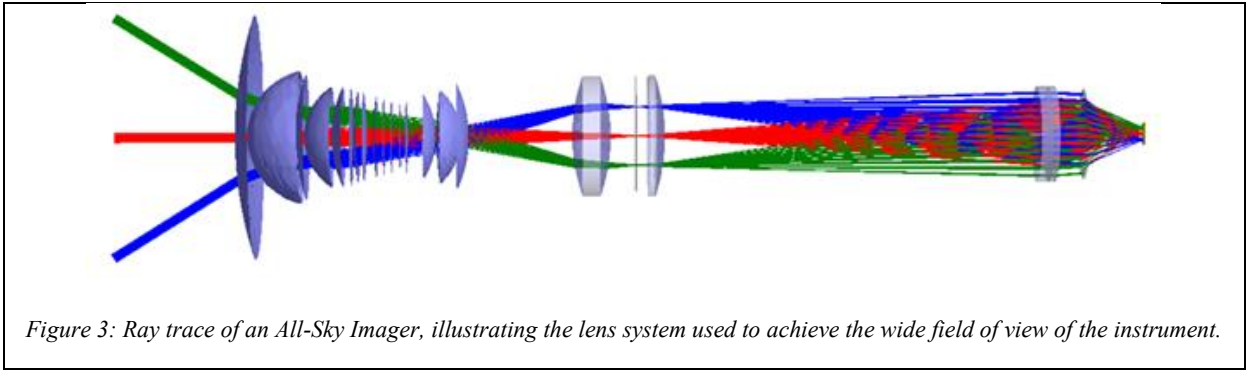

*Figure 3: Ray trace of an All-Sky Imager, illustrating the lens system used to achieve the wide field of view of the instrument.*

Second, the orientation of the image relative to True North needs to be determined. The ASI image does not have fiducial markers for establishing True North. Further, Figure 2 shows the ideal scenario where the imager is perfectly aligned with the zenith, that is, perpendicular to the horizontal plane. Real world coordinates of a point P at emission height H from above the Earth have a ground footprint at location L specified by a latitude and longitude.

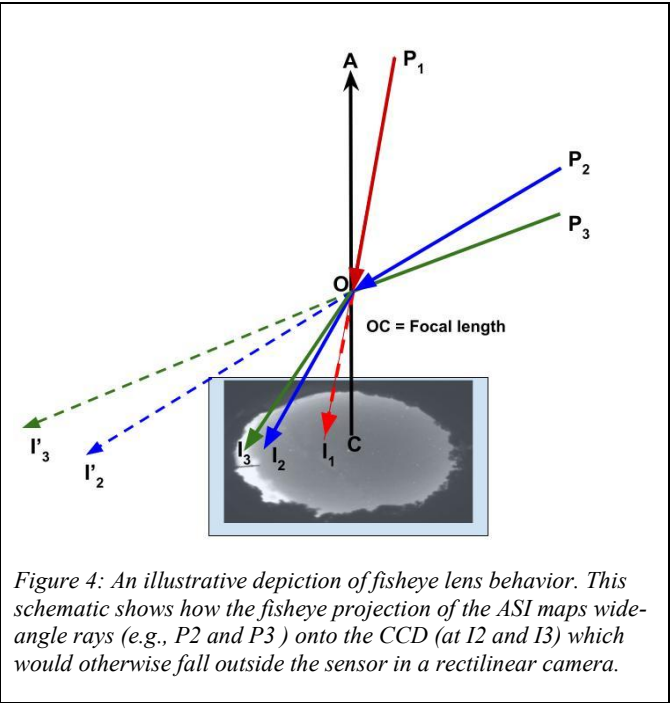

*Figure 4: An illustrative depiction of fisheye lens behavior. This schematic shows how the fisheye projection of the ASI maps wide-angle rays (e.g., P2 and P3 ) onto the CCD (at I2 and I3) which would otherwise fall outside the sensor in a rectilinear camera.*

In practice, perfect vertical alignment of the ASI is difficult to achieve, and installations may exhibit small roll, pitch, and yaw offsets. For example, Figure 5 shows the ASI in its ruggedized housing at the deployment site. Although every effort is made during deployment to ensure that the imager is level and oriented vertically (i.e., looking directly upward), perfect alignment is difficult to achieve. As a result, it is necessary to estimate not only the yaw angle (which determines the alignment with True North) but also the roll and pitch angles, which describe the camera's tilt relative to the vertical axis [Figure 6].

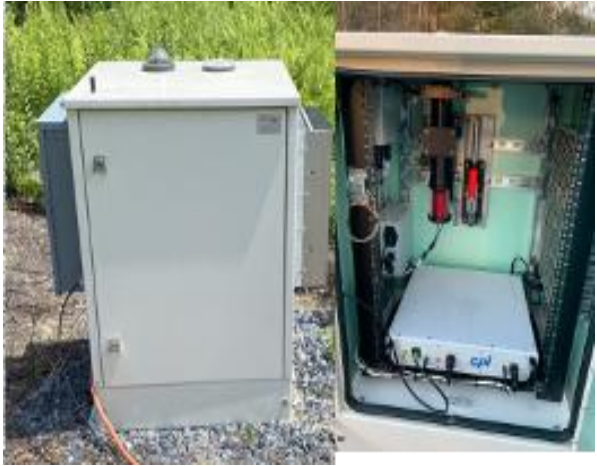

*Figure 5: ASI deployed at MIT Haystack, Westford, MA, shown installed in its ruggedized housing.*

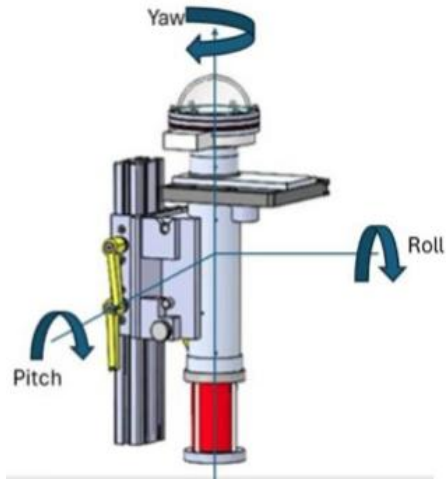

*Figure 6: Schematic diagram showing roll, pitch and yaw angles that determine the instrument pose.*

The combined effect of the fisheye lens behavior and need to estimate the camera orientation on the field together present a difficult challenge in determining the geo-reference map for the image data. Further, over time, mechanical drift, thermal expansion, or field handling can introduce small but significant orientation changes in the camera pose. As a result, ASI systems require continuous calibration, validation and error tracking to ensure consistent spatial accuracy throughout long deployments.

## 3.2 Related Work on Calibration Techniques

Several techniques have been developed for modeling the fisheye lens distortion of the ASI. The Scaramuzza model (Scaramuzza *et al.*, 2006) provides a flexible framework for calibrating omnidirectional cameras, including fisheye and central catadioptric systems with a polynomial that maps the radial distance of image points from the image center to the corresponding direction vectors in 3D space. The Kannala-Brandt model (Kannala *et al.*, 2006) is specifically designed for fisheye lenses and models the camera projection based on a spherical mapping between incoming 3D rays and image coordinates. It expresses the radial distortion as a trigonometric series of the incident angle, enabling precise modeling of the nonlinear projection behavior typical of fisheye optics.

Current state-of-the-art solutions involve using star maps to derive the fisheye lens distortion model for the ASI lens system and estimate the orientation of the ASI instrument. The THEMIS ASIs (Donovan *et al.,* 2006) use the sky map-based calibration files—mapping pixel coordinates to geographic locations at assumed auroral heights (e.g., 110 km). These sky maps must be periodically regenerated, especially following instrument relocation or adjustment, which can introduce geolocation drift unless regularly corrected. Furthermore, Donovan et al. (2006) describe the THEMIS ASI network architecture and emphasize the need for precise spatial registration across multiple sites to enable accurate, synchronized auroral imaging.

However, state-of-the-art methodology presents multiple problems in the delivery of real-time geo-referenced maps: manual calibration is time consuming, gradual calibration changes due to changing environmental conditions are solved by discrete recalibration exercises, the calibration results require frequent monitoring or may not detect errors from accidental changes in orientation due to an external bump which can otherwise go undetected for long periods.

*These operational realities underscore the importance of our fully automated Auto-Cal pipeline, which provides continuous recalibration and error tracking absent in traditional workflow of skymap-based systems.* The approach developed by Computational Physics, Inc. (CPI) meets this demand through robust automation, making available geo-referenced image data from All-Sky Imagers in near real-time together with error estimates. Auto-Cal performs nightly camera pose calibration using stellar references and lens distortion modeling, without the need for human intervention.

# 4 Methodology

Our solution uses a two-pronged approach to solve the spatial calibration of ASIs. The first step is to model the fisheye lens distortion of the imager to represent the relationship between the angular incident ray and radial displacement in the image. After correcting for lens distortion, the apparent circular motion of stars around the celestial pole is used to estimate its location in the image, by projecting the star tracks onto a 3D hemisphere based on the previously determined lens model. The apparent circular motion of stars is used to locate the celestial pole, from which an approximate yaw angle of the camera is computed. An optimization algorithm computes the roll, pitch and yaw angles by minimizing the error between observed and expected star positions.

In addition to producing traditional discrete lookup tables linking pixel indices to azimuth and elevation (and, by extension, geographic coordinates at a specific emission height), Auto-Cal provides methods for forward and inverse projections between altitude and azimuth angles and the pixel position with continuous geometry. Our system separately transforms unit hemisphere vectors (parameterized by altitude and azimuth) to the geometric image plane using the calibrated lens model, and from there to floating-point pixel coordinates. This yields a fully differentiable, continuous mapping that supports sub-pixel interpolation and more precise geolocation than typical lookup tables. Furthermore, the inverse transform—from any floating-point pixel location back to azimuth, elevation, and geographic location—is equally continuous and precise, enabling refined mapping of dynamic auroral structures.

## 4.1 Fisheye Lens Distortion Model

We use the Kannala-Brandt model in our calibration methodology to model the lens function of the ASI because it is specifically designed for fisheye lenses and offers high accuracy in modeling wide-angle distortions. Its spherical projection approach captures the nonlinear relationship between incident light angles and image coordinates, which is essential for precise geo-referencing of all-sky images. Additionally, the model is supported by the OpenCV library in Python, enabling seamless integration into our automated pipeline. This combination of theoretical suitability and practical implementation makes it ideal for continuous calibration in real-world ASI deployments.

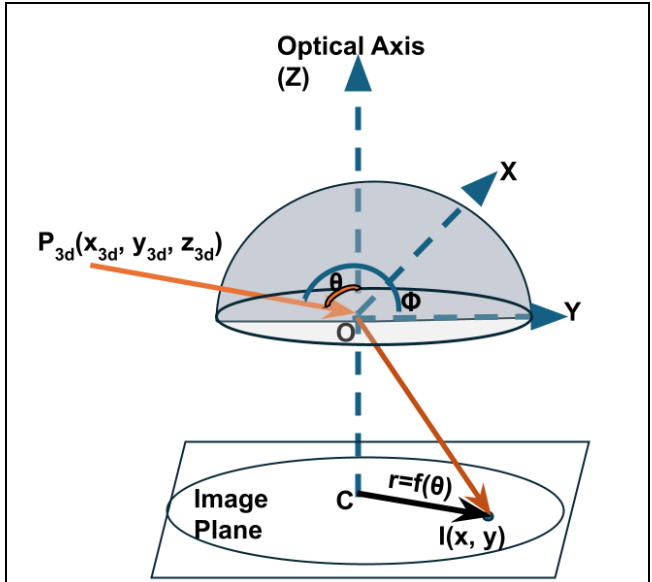

Figure 7 Geometric relationship between an object in 3D coordinates $P_{3d}(x_{3d}, y_{3d}, z_{3d})$ and its image coordinates $I(x,y)$ on the camera sensor plane using the Kannala-Brandt model. The 3D point makes a zenith angle $\theta$ with the optical axis and azimuth angle $\varphi$ in the horizontal plane with the y-axis.

The Kannala–Brandt model projects a 3D point onto the image plane by computing a distorted radial distance using a polynomial [Figure 7]:

$$r(\theta) = k_1\theta + k_2\theta^3 + k_3\theta^5 + k_4\theta^7 + k_5\theta^9 + \ldots,$$

where:

- $\theta$ is the angle between the 3D point and the optical axis. $\theta$ is computed from the direction of the incoming light ray in the camera frame, relative to the optical axis (usually the Z-axis).

- $k_1..k_4$ are distortion coefficients

- $r(\theta)$ is the **radial distance** from the optical center to the projected point on the image plane

### 4.1.1 Calibration Method

A series of checkerboard images are captured at varying angles and distances from the imager. [Figure 8]. Care is taken to ensure that the images are taken at varying depths and cover the full field of view. If the imager is equipped with a filter wheel, this procedure is repeated for each filter position to accurately calibrate the intrinsic camera parameters for that specific filter. This is necessary because filters may differ slightly in thickness or alignment, leading to small shifts in the optical path and effective image center.

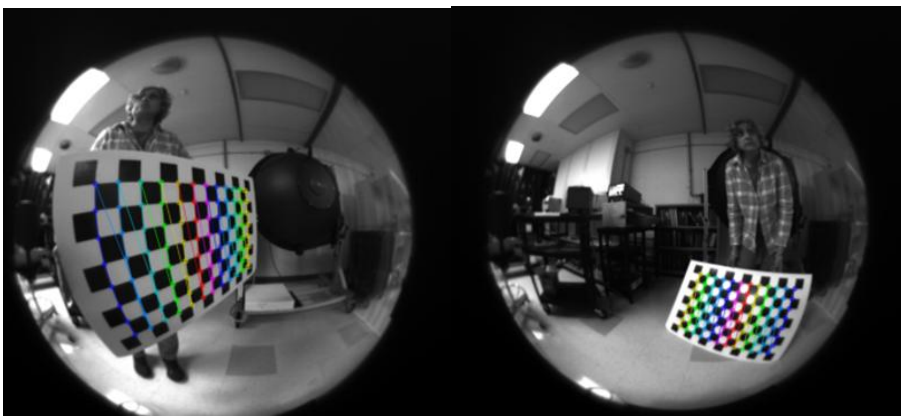

Figure 8: A series of images with a planar checkerboard pattern are acquired, such that they cover the field of view of the ASI image.

The checkerboard images corresponding to each filter are processed separately to derive a unique set of intrinsic parameters for that filter. The inner corners of the checkerboard are the control object points on a planar surface, whose projected corners in the image space are automatically detected using computer vision methods provided by OpenCV.

These corner positions, along with their corresponding control object points, are input to the fisheye calibration method of OpenCV, which is based on the Kannala-Brandt model. The calibration process estimates intrinsic camera parameters which include the optical center ($C_x$, $C_y$), focal lengths ($F_x$, $F_y$), and distortion coefficients ($k_1..k_4$). The radial distortion is modeled by the equation:

$$\theta_d = \theta(1 + k_1\theta^2 + k_2\theta^4 + k_3\theta^6 + k_4\theta^8) \qquad \text{Equation 1}$$

The following sections describe the function relationship by which the distortion angle $\theta_d$ maps the radial distance $r(\theta)$ to pixel coordinates of the image using the camera intrinsic parameters: the image center C and the focal length in pixels along x and y axis, $F_x$ and $F_y$. This is used to project the image points to 3D space and vice versa.

### 4.1.2 Mapping Between Image Coordinates and Object Points in Emission Layer

To compute the forward and inverse mapping of a point in the emission layer to its pixel location, we represent the 3D point in the emission layer by its altitude and azimuth coordinates on a unit hemisphere centered at the camera's optical center. This framework provides a natural and physically meaningful way to represent light originating from the upper atmosphere or celestial sources, such as auroral emissions, airglow, or stars. It enables both the forward projection of known atmospheric or astronomical features onto the image plane and the inverse mapping of image pixels back into spherical viewing directions. Notably, star tracks—used later in the calibration process—naturally follow great circles on the hemisphere due to the apparent motion of the celestial sphere, whereas their projections onto a 2D rectilinear plane form non-circular path. Representing viewing directions on the unit hemisphere provides a geometrically accurate and physically intuitive framework that corresponds directly to the camera's field of view, avoiding the fisheye distortion introduced at the image plane. This simplifies transformations related to camera pose, optical axis alignment, and image formation.

**Forward Projection of a point on 3-D unit hemisphere to Pixel Location**

In the forward projection process, we begin with a point on the 3D hemisphere defined by altitude and azimuth angular coordinates. During calibration, the positions of stars given in right ascension and declination can be readily converted to altitude and azimuth coordinates, provided the observer's location and elevation are known using well-established tools such as the Astropy library [Astropy Collaboration, 2018].

Given a 3D point on the unit hemisphere represented as (altitude, azimuth), its 3D coordinates on a unit hemisphere can be represented as $P_{3d}(x_{3d}, y_{3d}, z_{3d})$,

Where:

$z_{3d} = \sin(altitude)$

The horizontal projection, $r_u = \text{sqrt}(1 - z_{3d}^2)$, and

$x_{3d} = \cos(azimuth) * r_u$

$y_{3d} = \sin(azimuth) * r_u$

Then,

$\theta = \arccos(z_{3d})$ is the angle between the ray and optical axis, and

$\theta_d$ is computed using the distortion polynomial given in $\boldsymbol{\theta_d = \theta(1 + k_1\theta^2 + k_2\theta^4 + k_3\theta^6 + k_4\theta^8)}$     Equation 1

The x and y components of the horizontal projection of the unit vector are normalized as follows:

$x_u = \frac{x_{3d}}{r_u}, y_u = \frac{y_{3d}}{r_u}$

On the image plane, the distorted normalized geometric coordinates ($x_d$, $y_d$) are computed as:

$x_d = x_u * \theta_d$ and

$y_d = y_u * \theta_d$

Finally, the image coordinates with pixel locations I(x, y) are computed as:

$x = f_x * x_d + c_x,$
$y = f_y * y_d + c_y$

**Reverse Projection of a Pixel on Image Plane to 3-D Unit Hemisphere**

For the inverse mapping—from pixel coordinates to the angular direction of incoming light in altitude and azimuth coordinates —the radial distance **r** of the pixel I(x,y) from the optical center ($C_x$, $C_y$) is used to estimate the zenith angle θ between the ray and the optical axis. This relationship is modeled using a polynomial of the form below, which are obtained polynomial curve fitting of the forward Kannala–Brandt distortion model:

$\theta = P_0 + P_1 r + P_1 r^2 + P_2 r^3 + P_3 r^4$ and

Altitude angle $= \frac{\pi}{2} - \theta$

The azimuth angle is simply $\arctan(\frac{y}{x})$

This avoids runtime root-solving or iterative inversion by using a fast, direct evaluation of the fitted polynomial to approximate θ from r. This trade-off provides a good balance between speed and accuracy in continuous calibration workflows.

## 4.2  Estimation of Camera Pose with Star Tracks

After correcting for lens distortion, the next essential step is to determine the orientation of the All-Sky Imager (ASI) with respect to geographic (True) North. This orientation is critical for accurately computing the ground footprint of observed emissions at known altitudes, allowing their 3D positions to be expressed in geographic coordinates (latitude, longitude).

The camera pose of the All-Sky Imager (ASI) is determined through a two-stage process:
 (1) **Estimation of the celestial pole** from observed star tracks, and
 (2) **Alignment of observed stars with a celestial catalog** using optimization.

To estimate the ASI's full pose (yaw, pitch, and roll), the system uses sequences of night-sky images to detect and track the apparent motion of stars. These tracks are projected onto a 3D hemisphere using the previously determined lens model, which eliminates the distortion due to the ASI lens system. The observed circular motion of stars, caused by Earth's rotation, is used to identify the position of the celestial pole. This serves as a reference point from which the orientation angles of the ASI are inferred.

An optimization algorithm is then employed to refine the initial pose estimates. The algorithm minimizes the discrepancy between observed star positions and those predicted by astronomical models, iteratively adjusting the roll, pitch, and yaw angles to achieve the best fit. This process ensures an accurate determination of the ASI's orientation, which is essential for the precise geolocation of sky-based emissions.

### 4.2.1  Identifying the Celestial Pole in Pixel Coordinates

The images acquired by the imager are sorted in time and the filter used in the observation. They are then pre-processed using a median filter to eliminate hot pixels.

**Star Detection**

Star-like features are identified in ASI image sub-regions using a Laplacian of Gaussian (LoG) blob detection algorithm, implemented in OpenCV. The method detects bright, approximately circular features within a defined scale range (2-5 pixels in sigma).

This method detects all bright objects that have star-like features in the image, including planets and reflections caused by moonlight or environmental contaminations. For any given night, planets track circular path around the celestial pole. Reflections on the imager dome due to the moon or other environmental conditions do not circle the celestial pole. These are eliminated in the next stage of processing, described later in the document. Further, not all stars are detected in every image as some stars may be occluded due to clouds or the moon. The pose of the instrument is not known at this stage, so stars are not labeled or identified.

**Identification of Star Tracks**

To reliably identify stars in the All-Sky Imager (ASI) data, the algorithm compares features identified as stars across successive images taken at regular intervals. The cadence of images acquired with a specific optical filter is determined by the data acquisition plan of the acquisition program. From experience, it was determined that the star tracking method works with image cadences intervals up to 15 minutes. In our operations, the typical image cadence is approximately 3.5 minutes. Not all bright star-like features detected in a single image frame by the start detection algorithm are actual stars; therefore, a feature is only accepted as a star if it lies within the field of view of the imager and a corresponding detection exists in the next frame within a small displacement threshold (less than 15 pixels). Stars detected consistently across successive frames are recorded in a displacement list with an entry that captures the star's pixel location and displacement in the x and y directions, along with the timestamp and time interval. As illustrated in Figure 9 and Figure 10, the resulting star tracks form smooth, curved paths that closely follow expected stellar trajectories. Red dots in the figure represent individual star positions detected across frames, while the arcs highlight consistent stellar motion over time.

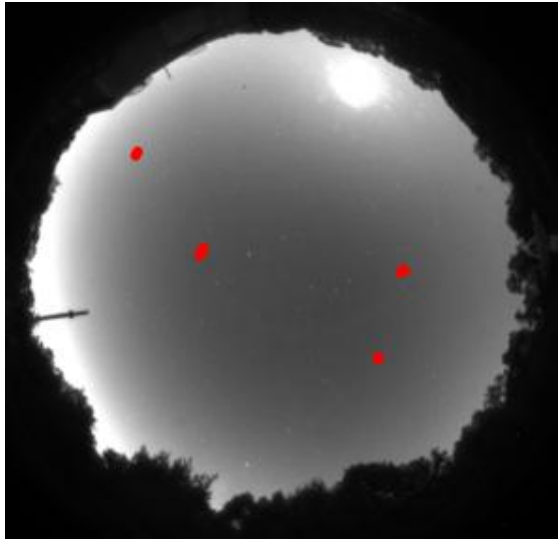

*Figure 9: Tracked star motion are overlaid on ASI images taken on 06-02-2025, a few images into the beginning of data acquisition. Red dots indicate individual star positions detected across consecutive frames. This shows the initial part of the star tracks.*

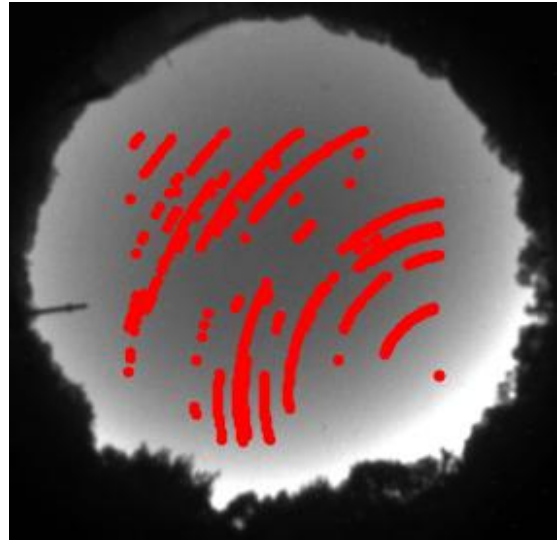

*Figure 10: Tracked star motion are overlaid on ASI images taken on 06-02-2025, towards the end of data acquisition. The curved pattern represents consistent star movement due to Earth's rotation. The star tracks are computed by matching blobs between image pairs, with motion vectors filtered by displacement and spatial consistency within the field of view. Some of the tracks are very short as they may have been occluded by passing clouds.*

These star displacement lists are accumulated over time to form star tracks—contiguous sequences of detections that represent the apparent motion of stars due to Earth's rotation. The set of detected star displacements are represented as flow vectors with start and end positions of a detected star movement [Figure 11]. The flow vectors are grouped into distinct star tracks, representing continuous motion of individual stars across the ASI field of view as follows: For each vector in the input sequence, the algorithm computes the motion direction and magnitude. It then iteratively samples points along the vector's path and checks a 2D spatial map to determine if the path intersects any existing arc. If so, the displacement is assigned to the corresponding arc ID; otherwise, a new arc ID is created. The arc ID is then propagated along the path of the flow vector by updating the 2D spatial map. This approach ensures that spatially overlapping or adjacent motions are grouped into coherent star tracks, facilitating the reconstruction of stellar trajectories over time. The output is an updated table with assigned arc identifiers and a mapshowing the spatial distribution of arcs across the image [Figure 12].

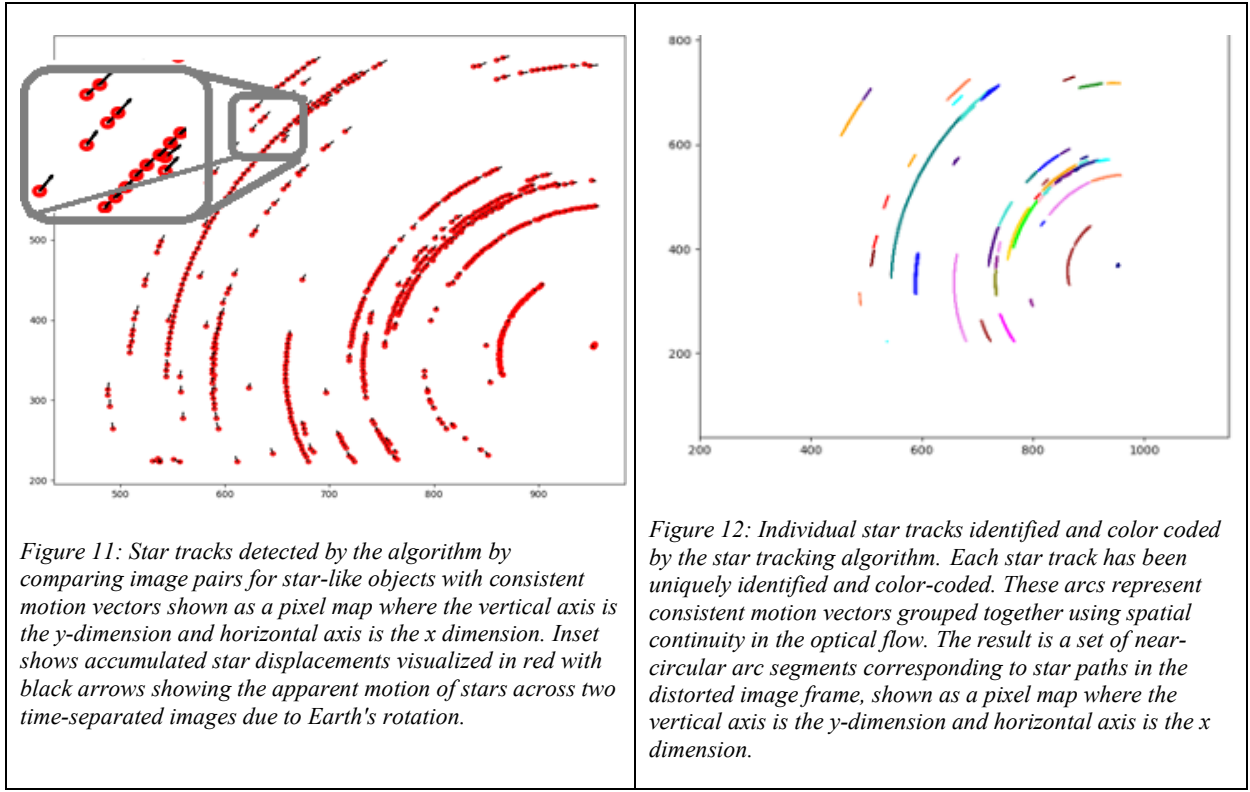

Figure 11: Star tracks detected by the algorithm by comparing image pairs for star-like objects with consistent motion vectors shown as a pixel map where the vertical axis is the y-dimension and horizontal axis is the x dimension. Inset shows accumulated star displacements visualized in red with black arrows showing the apparent motion of stars across two time-separated images due to Earth's rotation.

Figure 12: Individual star tracks identified and color coded by the star tracking algorithm. Each star track has been uniquely identified and color-coded. These arcs represent consistent motion vectors grouped together using spatial continuity in the optical flow. The result is a set of near-circular arc segments corresponding to star paths in the distorted image frame, shown as a pixel map where the vertical axis is the y-dimension and horizontal axis is the x dimension.

**Estimation of Celestial Pole using Projection of Star Tracks on to 3D Hemisphere.**

Next, using the known camera calibration parameters, the pixel coordinates of all arc-associated points are unwarped and projected onto a 3D unit hemisphere. This projection step compensates for lens distortion and maps each star detection to a unit vector representing its direction in the celestial hemisphere [Figure 13].

Each valid star track (i.e. one that is consistently identified over a duration of at least about half hour) is then processed to estimate the center of its circular path on the unit hemisphere. These circular arcs result from the apparent circular motion of stars around the celestial pole. The center of each arc is estimated using a least-squares optimization that minimizes variance in geodesic distance from a candidate center to the arc points—effectively fitting a small circle on the sphere.

The estimated 3D center of the arc is then projected back into image (CCD) coordinates. These estimated centers from multiple arcs are aggregated, and their median position is computed to yield a robust estimate of the celestial pole in image space. The system supports deployment in both hemispheres by identifying the appropriate celestial pole (North or South) based on star trajectories which appear to circle the celestial pole in both hemispheres.

This approach ensures that lens distortion is eliminated through 3D modeling, while preserving the circular nature of star tracks that encode celestial motion.

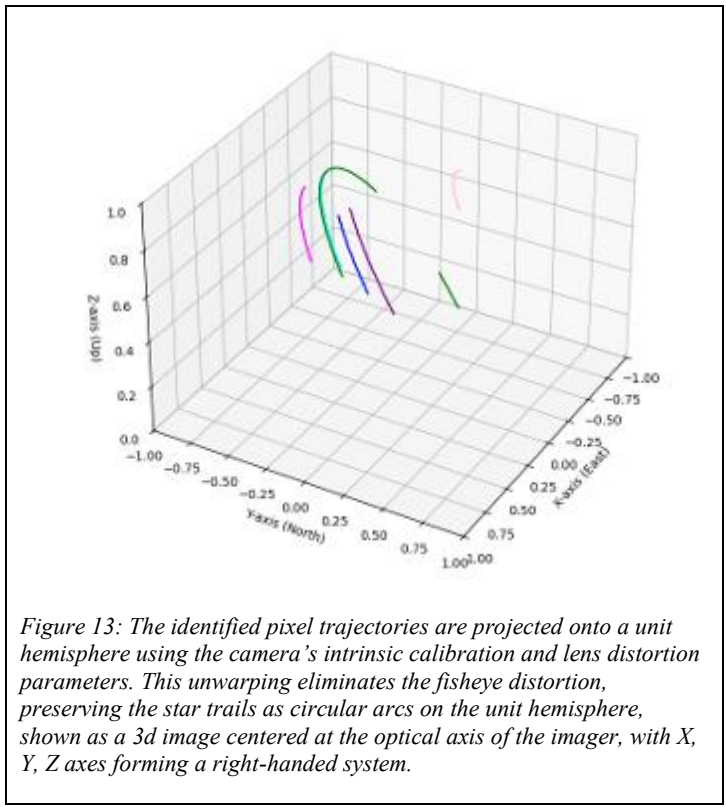

*Figure 13: The identified pixel trajectories are projected onto a unit hemisphere using the camera's intrinsic calibration and lens distortion parameters. This unwarping eliminates the fisheye distortion, preserving the star trails as circular arcs on the unit hemisphere, shown as a 3d image centered at the optical axis of the imager, with X, Y, Z axes forming a right-handed system.*

**Robustness to Various Observing Conditions**

The calibration method ensures robustness against various conditions such as cloud cover and moonlight by using the median of the celestial pole estimates of all the star tracks. This ensures that outliers such as specular reflections of light due to the Moon, which have "star-like" tracks, are ignored. Poor observing conditions result in a higher standard deviation of the estimated celestial pole position. Because a large standard deviation indicates uncertainty in the estimated celestial pole position, a threshold of half the focal length is used as a heuristic for determining where it is possible to compute the initial yaw angle estimate, a necessary step in the computation of the roll, pitch and yaw rotation angles.

On nights when observing conditions are not ideal for a full calibration of the rotation angles but clear skies are available for part of the night, the system uses the most recently computed rotation angles from past calibrations to estimate errors. As a result, error estimates are provided whenever possible. However, if cloud cover or poor conditions persist throughout the entire image set and no stars are visible, neither calibration nor error estimates can be generated. This approach ensures that users have the necessary information to assess the reliability of the data.

### 4.2.2 Derivation of Camera Pose: Roll, Pitch and Yaw Angles

To refine the full 3D orientation of the camera (roll, pitch, and yaw), a set of bright reference stars is used. The software maintains a catalog of known bright stars indexed by latitude. Using the **Astropy** library, the precise Altitude–Azimuth coordinates of these catalog stars (e.g., Polaris, Vega, Sirius) are computed for the ASI's geographic location and image timestamp. These coordinates are converted to 3D unit vectors on the celestial hemisphere, representing the **true directions** to the stars.

An initial yaw estimate is derived from the computed direction to the celestial pole. This estimate, combined with the calibrated lens distortion model, is used to forward-project the true 3D star directions into approximate 2D pixel coordinates on the image plane. These projections define the search regions where the actual star images are expected to appear [Figure 14].

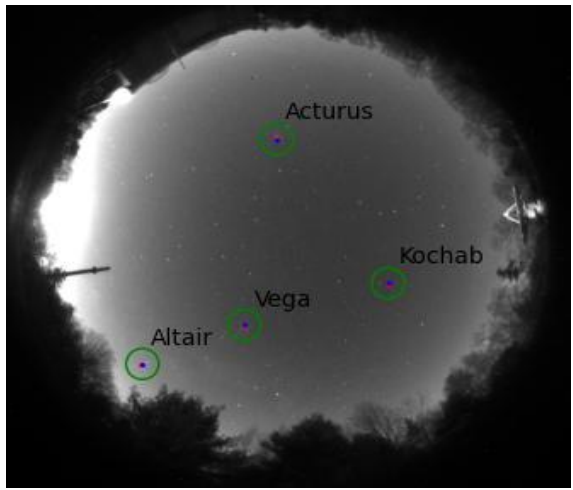

*Figure 14: Star overlap after optimization. The red dots represent expected star positions post-optimization, taking into consideration the optimized camera pose and applying the distortion model. The blue dots represent the actual star positions detected in the search region shown as green circles.*

A blob detection algorithm is applied in these regions to identify the observed star locations in the ASI image. The detected pixel positions are then unwarped and mapped back onto the unit hemisphere to yield observed 3D direction vectors.

A least-squares optimization is performed to compute the rotation matrix that best aligns the observed vectors with the known star directions. This rotation matrix captures the camera's orientation relative to the geographic frame (true north and zenith). It is subsequently applied to geo-reference the ASI imagery in downstream analysis.

# 5 System Integration and Automation

The calibration system is embedded within CPI's data processing pipeline. It recalibrates automatically each night, adapts to environmental changes, and provides error estimates. This ensures high reliability and transparency in data quality assessments. Importantly, the system can self-correct after being physically moved or reoriented, making it ideal for remote deployments.

When cloud cover prevents estimation of the celestial pole or camera pose, the system flags the data as uncalibrated. The absence of calibration on any given night is handled by using the continuous data processing handling of the calibration estimates, which replaces the "current" best estimate only when better estimates are available for the night.

# 6 Results & Validation

The technique has been successfully implemented and tested for the ASI deployed at MIT Haystack Observatory since May 2025. It performs reliably under varying conditions, including partially cloudy nights and moonlit skies. Visual results—such as tracked star paths and post-calibration error maps—demonstrate quantifiable accuracy in sky mapping. The system's output includes nightly error estimates, enabling data users to assess the reliability of the calibration for individual dates [Figure 15]. Currently, the error estimates are reported in pixel coordinates as Mean Euclidean Error (MEE), that is, the mean Euclidean distance between the expected and actual star positions. However, the true ground distance represented by each pixel varies across the image because the ground footprint of the pixel depends on the angle of incidence. Therefore, the same error in pixel units may correspond to different physical distances on the georeferenced footprint.

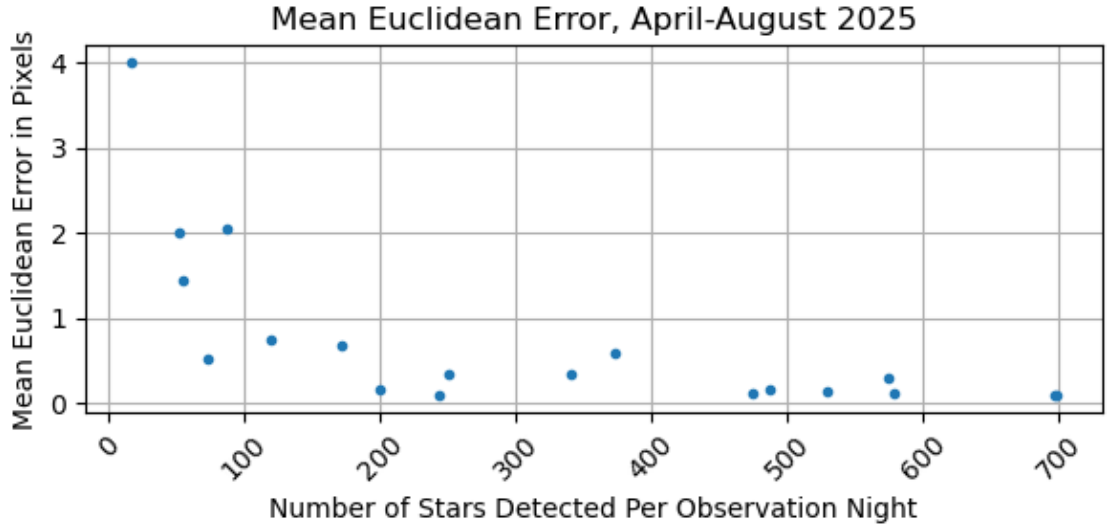

*Figure 15: Mean Euclidean Error in pixel dimensions, measured as the average Euclidean distance between expected and actual star positions. The Y-axis is the Mean Euclidean Error, and the X-axis is the number of stars identified in the images for the night. Fewer stars imply the sky had either too much contamination from moonlight or was cloudy.*

The estimation of the camera pose by computing the celestial pole and star tracking is computed every night. Currently the calibration is run with one night of data acquisition. This methodology will be adapted to real-time workflows by fitting it into the CPI's general data processing pipeline, where the real-time data is calibrated with a best-estimate, and a refined, more precise data stream is generated as more data becomes available.

## 6.1 Runtime Performance

Fisheye lens calibration is conducted offline during instrument assembly, ensuring it does not impact real-time performance.

While the current implementation prioritizes accuracy, robustness, and automation, it has not yet been optimized for runtime performance. As such, calibration runtime may vary depending on system hardware and environmental conditions. Preliminary benchmarks indicate that the nightly calibration process completes in under 2.5 minutes on standard computing hardware. Future work will focus on optimizing key modules for parallel processing such as star detection to support real-time deployments.

| Stage | Runtime (s) |
| --- | --- |
| Estimation of celestial pole | 136.6 |
| Estimation camera pose (roll, pitch and yaw angles) | 3.4 |

*Table 1: Runtime Performance of calibration process*

# 7 Impact and Applications

This work significantly advances the capabilities of ASI networks. It reduces the burden of manual calibration, enhances spatial accuracy, and supports deployment in remote areas. In CPI's experience, spatial calibration of ASIs using traditional manual calibration takes a few hours. More importantly, it is carried out only after significant errors are apparent, or when the instrument is redeployed after an upgrade. Auto-Cal provides the capability to automatically calibrate the instrument every night. Further error estimates are produced along with the calibration results, thus enabling the user to determine the trustworthiness of the imager data.

Installation of a second imager is planned for Easton, Maine, enabling inter-site feature tracking across larger geographic regions. The project establishes a scalable foundation for long-term geophysical observation networks.

# 8 Discussions and Limitations

Auto-Cal encounters a fundamental geometric limitation near the geographic poles. The algorithm relies on star tracking and estimation of the celestial pole, which degenerates at exactly 90° latitude due to the singularity of coordinate transformations at that point. Although the system performs reliably at high latitudes, calibration exactly at the North or South Pole remains an open challenge. Future work will focus on developing pole-specific strategies, such as advanced astrometry-based solutions using star patterns, to extend usability to true polar deployments.

# 9 Conclusion

This project delivers a practical, automated method for spatial calibration of ASIs. By unifying star tracking and distortion correction, this system enables reliable, high-accuracy geophysical observations with minimal operator intervention. The foundation laid by this work is instrumental for the expansion of global all-sky imaging networks and real-time monitoring of upper atmospheric dynamics.

# 10 Acknowledgments